\documentclass[12pt,a4paper]{article}
\usepackage[utf8x]{inputenc}
\usepackage{ucs}
\usepackage[left=2.5cm,top=2.5cm,right=2.5cm,bottom=2.5cm]{geometry}
\usepackage{authblk}
\usepackage[authoryear]{natbib}
\usepackage{amsmath,amsfonts,bm,amssymb,latexsym,amsthm}
\usepackage{graphicx, graphics}
\usepackage{multirow}
\usepackage[authoryear]{natbib}
\usepackage{microtype}
\usepackage[none]{hyphenat}
\usepackage{setspace} 
\usepackage{array}
\usepackage{amsfonts}   
\usepackage[bottom]{footmisc}
\usepackage{color}
\usepackage{url}
\usepackage{float}
\usepackage{rotating}

\usepackage{soul} 
\usepackage{todonotes} 

\usepackage{placeins}
\usepackage{multicol}
\usepackage{multirow}

\usepackage{morefloats}
\usepackage{booktabs}
\usepackage{multirow}
\usepackage[para,online,flushleft]{threeparttable}
\usepackage[justification=justified,font=small]{caption}

\usepackage{pifont}

\usepackage{subcaption} 

\usepackage{svg}

\usepackage{comment}

\newcolumntype{C}[1]{>{\centering\let\newline\\\arraybackslash\hspace{0pt}}m{#1}}

\usepackage{hyperref}
\hypersetup{colorlinks, breaklinks, linkcolor=[rgb]{0,0.3,1}, citecolor=[rgb]{0,0.3,1}, urlcolor=[rgb]{0,0.3,1} }
\usepackage{longtable}

\begin{document}

\title{Being at the core: firm product specialisation}

\author[1,2,*]{Filippo Bontadini}
\author[3]{Mercedes Campi}
\author[4,5]{Marco Due\~nas}

\affil[1]{\small LUISS University, Italy}
\affil[2]{\small SPRU - University of Sussex, United Kingdom}
\affil[3]{\small CONICET - Universidad de Buenos Aires, Instituto Interdisciplinario de Econom\'ia Pol\'itica de Buenos Aires (IIEP), Argentina}
\affil[4]{\small IMT School for Advanced Studies, Lucca, Italy}
\affil[5]{\small Institute of Economics - Sant'Anna School of Advanced Studies, Pisa, Italy}
\affil[*]{Corresponding author}

\date{}

\maketitle
\vspace{-40pt}

\begin{abstract} 
\noindent 
We propose a novel measure to investigate firms' product specialisation: \textit{product coreness}, that captures the centrality of exported products within the firm's export basket. We study product coreness using firm-product level data between 2018 and 2020 for Colombia, Ecuador, and Peru. Three main findings emerge from our analysis. First, the composition of firms' export baskets changes relatively little from one year to the other, and products far from the firm's core competencies, with low coreness, are more likely to be dropped. Second, higher coreness is associated with larger export flows at the firm level. Third, such firm-level patterns also have implications at the aggregate level: products that are, on average, exported with higher coreness have higher export flows at the country level, which holds across all levels of product complexity. Therefore, the paper shows that how closely a product fits within a firm's capabilities is important for economic performance at both the firm and country level. We explore these issues within an econometric framework, finding robust evidence both across our three countries and for each country separately.
\end{abstract}



\clearpage
\newpage

\section{Introduction}

There is broad evidence that what countries export is related to their level of development \citep{hausmann2007you}. Countries grow as they acquire capabilities and learn to combine them to diversify exports \citep{hidalgo2007, neffke2011regions, felipe2012product, zaccaria2014}. There is evidence of similar processes at the firm level: firms that export coherent baskets of products have better performance \citep{teece1994, patel1997, bottazzi2010coherence, dosi2022coherence}. 

In this paper, we focus on how close the products of the firms' export baskets are to their capabilities and how this impacts their export performance. We use customs level data for Colombia, Ecuador, and Peru for 2018--2020, allowing us to test these ideas in different contexts, including the COVID--19 crisis, which has led to an unprecedented sanitary and economic crisis. 

The evidence of the COVID--19 crisis on firms in developing countries suggests that small businesses have been particularly affected due to financial constraints \citep{bartik2020impact} as well as supply chain disruptions and demand-side shocks \citep{shafi2020impact}.Evidence of the impact on trade in Latin American countries, while still scarce, has begun to emerge. For Colombia, \cite{benguria2021collpase} studies how exporting firms responded to the trade collapse that followed the COVID--19 outbreak. He finds that despite a large and sudden drop in the number of exporters, most of the decline in exports is accounted for by intensive margin adjustments. While trade flows have decreased considerably during the COVID--19 crisis, there is likely to be heterogeneity across different types of firms and products. Also, for Colombia, \cite{campi2022bid} show that belonging to an export cluster does not automatically lead to higher firm resilience and that there are differences in how firms react to a crisis within clusters. Larger and more diversified firms have higher chances of surviving. Furthermore, this probability is even higher for products belonging to the firm's cluster, which suggests that the products that are closer to firms' main capabilities are more resilient.

The evidence above highlights that the pandemic likely affected firms differently, depending on what products they export and how efficient they are at exporting them. These issues are at the centre of two key strands of literature that are particularly relevant to our study.

First, the thriving literature on multi--product firms has highlighted that firms enter export markets with their most efficient product, expand their scope with those products that face lower marginal costs and that sales are skewed towards the firm's better-performing products \citep{eckel_neary2010, bernard2010, bernard2011, mayer2014}. We link this with another long-standing strand of research that has viewed firms as a bundle of resources and capabilities and has explored the importance of the coherence of such a bundle \citep{penrose1959, teece1994}.

Second, a more recent strand of literature has emerged from this capabilities framework, exploring the relationship among products in terms of proximity within a product space \citep{hidalgo2007} and as a proxy for firms' capabilities \citep{dosi2017firmsknow, bruno2018colombian, fontagne2018exporters, campi2022bid}. At the country level, the product space literature has put forward a well-established measure of product complexity, finding more complex products to be associated with higher growth rates and economic development \citep{hidalgo2007, servedio2018new}. At the firm level, \cite{bruno2018colombian} use data on Colombian firms and show that firms specialise in subsets of products that lie close to each other in the product space and \cite{campi2022bid} used community detection tools to derive clusters of products. Furthermore, and particularly relevant to our purpose here, \cite{dosi2022coherence} have looked at the degree of coherence of the product baskets at the firm level, i.e., similarity within a firm's product portfolio, and found it to be associated with higher firm profitability.

This paper contributes to these strands of literature by studying the relationship between trade performance and product characteristics, especially their complexity and how close they are to other products exported by the same firm. Our key conjecture is that products that fit better within a firm's export portfolio will be associated with larger export flows and that this is likely to change across different levels of product complexity.

To test this, we devise a novel measure of \textit{coreness} that captures the proximity of exported products within the firm's export basket. We argue that this measure is useful to assess the position that each product occupies within a firm's export basket and, as a result of this, its trade performance. We show this with data on firm-product exports for Colombia, Ecuador and Peru between 2018 and 2020, therefore, including the first COVID--19 outbreak in our analysis. 

Our analysis yields three main results. First, firms tend to export products close to their core capabilities and these are less likely to be dropped from their export portfolio. Second, firms export products with high coreness in larger quantities. We find this relationship to be very robust, also when controlling for firms' diversification and whether the product is part of the firm's typical product mix \citep{fontagne2018exporters}. Furthermore, the positive relationship between coreness and export volumes persists across products with different levels of complexity. Third, these firm-level dynamics have bearing on how a country's export portfolio is likely to evolve, which in turn has implications for economic development \citep{hidalgo2007, hausmann2007you}. If certain products are exported with higher coreness--i.e., they are on average closer to the capabilities of firms that export them--countries' export portfolio is likely to pivot towards them. We test this conjecture, and its implications in terms of countries' export portfolios and find that products that are on average exported with higher coreness are exported in larger volumes.

We provide robust evidence that not only what firms export \citep{hausmann2007you, hausmann2011network} and the composition of the firms' export basket matter \citep{dosi2022coherence} matter, but that also the closeness of each product with respect to the firm's export baskets can reveal valuable information about their performance. 

The remainder of this study is structured as follows. Section~\ref{literature} discusses the relevant strands of literature and the conceptual framework of our study. Section~\ref{Data} presents the data and methods, including some key facts on the intensive and extensive margins of trade; our new measure of coreness. Section~\ref{results} illustrates the econometric application of this measure to see how firms have fared during the pandemic and what this means for country-level export specialisation. Finally, Section~\ref{conclusions} concludes and provides policy implications. 




\section{Related literature and contribution}
\label{literature}

A well-established empirical fact is that international trade--and arguably the economy in general--is dominated by large, multi-product firms that successfully export different goods to many destinations. This empirical regularity has spurred interest in the international trade literature, which has now put forward several models with heterogeneous (multi-product) firms. \cite{bernard2010, bernard2011} combine firm- and product-specific features to estimate sales differences across a firm's products. \cite{eckel_neary2010} use a core-competence model that assumes that firms manufacture multiple products using a flexible production technology: the efficiency in supplying less-successful products declines for products farther from firms' core competencies. Alternatively, \cite{mayer2014} assume firm-product-specific marginal costs that increase in scope together with non-CES preferences, which generates the prediction that firm sales will be more skewed towards their core competences in more competitive destination markets. \cite{montinari2021} highlighted the importance of investment in R\&D as a driver of the process of adapting the portfolio of companies in international markets.

This stream of work focuses on firms' production costs and efficiency as the main constrain to diversification. The overall conclusion is that firms enter foreign markets with their most efficient product and expand their scope starting with those products that face lower marginal costs and that sales are skewed towards the firm's better-performing products. 

Building on this, \cite{fontagne2018exporters} show that there exists great variability in the product mix that firms export across destinations. The authors argue that this is hard to reconcile with the idea of a product hierarchy that is stable across product markets, being driven by production efficiency \citep{eckel_neary2010, arkolakis2021extensive} or product quality \citep{eckel2015multi, manova2017multi}. In fact, they find bundles of products that firms export together, which they refer to as a firm's typical product vector (TPV). In this framework, \cite{fontagne2018exporters} put forward that product complementarities and technological relatedness are factors driving the composition of firms' typical product vectors.

This idea resonates with another well-established literature that has taken the view that firms' diversification is shaped by non-tradable, often intangible, resources -- which this literature refers to as capabilities. This view is well rooted in the seminal contribution of \cite{penrose1959}. In this framework, the firm is a set of resources and intangible capabilities, organised in an administrative framework. A firm's final production, at a given moment, represents one of several ways in which it can combine its resources.

The creation of a new product depends greatly on the possibility of enlarging the set of capabilities available to the firm. \cite{penrose1959} argues that firms accumulate capabilities with different uses. This conjecture has found some empirical support in more recent contributions, arguing that firms can maintain technological capabilities in several fields with a knowledge base that is wider than what their production would suggest \citep{brusoni2001, dosi2017firmsknow}. This excess of resources -- in other words, the fact that firms know more than they produce -- is a key factor in explaining product diversification. 


There are, however, limits to economies of scope and diversification.\footnote{Economies of scope occur when the joint cost of producing two different goods is smaller than producing them separately \citep{panzar1981}.} \cite{teece1994} show that existing capabilities constrain diversification opportunities of firms and that product portfolios are not random. The likelihood of a product being successfully included in a firm's production portfolio depends on whether the product is linked to the firm's core activities. From a firm perspective, it then becomes important to look at the coherence of its production, i.e., whether there exist certain technological and market characteristics common to each line of business or product \citep{teece1994}.

The relationship among products in a firm's portfolio and her ability to diversify has drawn growing attention. There is now a considerable body of evidence associating measures of coherence to higher stock market values \citep{nesta2006firm}, firm performance \citep{pugliese2019coherent}, survival as incumbent \citep{valvano2003diversification}, and growth and profitability \citep{piscitello2004corporate}.


Firms, therefore, diversify by branching out into products closer to their existing capabilities. \cite{bottazzi2006gibrat} show that this is, in turn, related to firms' size: larger firms will have larger bundles of capabilities, providing them with a broader base to branch out of and making them more likely to diversify. Specifically, this branching mechanism predicts that there exists an exponential relationship between a firm's size and the number of exported products. More recently, \cite{campi2018diversification}, looking at Chinese firms, confirmed this exponential relationship, giving further evidence to the branching mechanism linked to firms' size. 


Based on the literature discussed above, three observations are in order. First, there is a broad agreement that multi-product firms do not expand randomly, but follow product hierarchies. The scholarship has identified different drivers for this empirical result. On the one hand, the trade literature on multi-product firms has emphasised product quality and efficiency as the main drivers, on the other hand, the literature on firm capabilities has stressed the role of technological complementarities and product relatedness. Second, within this latter stream of research firms' capabilities enable, but also constrain the diversification process. Because of this, firms' production portfolios are informative of their capabilities. Third, the way in which firms diversify is relevant to their performance. If firms exploit their economies of scope, then the diversification process will be coherent with the firm's capabilities and likely be conducive to better economic performance. 

The literature has also explored these issues at the country level, using data on international trade \cite[see:][]{hidalgo2007, caldarelli2012, zaccaria2014}. A review of this vast stream of work is beyond our scope. However, the key contribution of this literature is that by relying on co-occurence in countries' export portfolios, a product space emerges in which products are mapped with respect to one another, and proximity is interpreted in terms of similarity of capability requirements \citep{hausmann2011network}. 

This country-level literature is that it has led to different measures of a product's complexity \citep{hidalgo2007, zaccaria2014}. In a nutshell, this approach considers that a product is complex if only some countries that are highly diversified (i.e., have a broad set of capabilities) export it (i.e., it requires capabilities that are not easily found among countries). Complex products, therefore, require a broad set of rare capabilities and are therefore harder to export. The literature on economic complexity has also made some significant contributions to the debate around economic specialisation and development and whether certain specialisation patterns are more beneficial than others. In their seminal contribution, \cite{hidalgo2007} find specialisation in complex exports to be associated with economic growth. As a result, economic complexity has gained significant attention in the policy debate around structural change, especially for developing countries \citep[see, e.g.,][]{britto2019great, ferraz2021linking}.

\subsection{Contribution and research questions}
\label{RQ}
Our contribution continues along the lines of thought detailed above, both at the firm and country level. Our approach is grounded in the firms' capabilities literature rather than production efficiency put forward in the trade literature on multi-product firms. Accordingly, we analyse the bipartite firm-product network to account for firms' diversification possibilities without having a detailed knowledge of production costs. This methodology can provide valuable information on firms' core capabilities, which can be used to understand their relationship with trade performance.

While the idea of coherence of firms' capabilities has been explored \citep{teece1994, patel1997, bottazzi2010coherence, dosi2022coherence}, we focus specifically on the importance of the relationship between a given product and the rest of a firm's export portfolio and ask whether products that are at the core of firms' exports exhibit better trade performance than those that are less close to firms' capabilities.
Importantly, this product-firm analysis allows us to also draw country-level implications. The starting point for this is twofold. First, our firm-level analysis shows that the coreness of a product within firms' export portfolios is important for export performance. Second, the literature on product complexity has shown that complex products are conducive to economic growth \citep{hidalgo2007, neffke2011regions, felipe2012product}. By bringing these two notions together, it becomes important to explore whether the average coreness with which products are exported by firms is related to trade volumes at the aggregate level and how this relationship varies across products with different complexity levels.

We test these ideas with highly disaggregated trade data that cover the COVID--19 pandemic, which we present in the next section, along with some stylised facts on firms' export diversification patterns.

\section{Data and methods}
\label{Data}

We use trade data for three Latin American countries: Colombia, from the Colombian Customs Office (DIAN); and Ecuador and Peru, from \cite{legis}. We consider monthly export transactions for 2018--2020 at the firm-product levels, with products described by the Harmonized System (HS) codes. For each transaction, we use the exporter's tax identification number (NIT), the month, a 6-digit HS code, the country of destination, and the free on board value in US dollars. We remove all transactions related to re-exports of products elaborated in other countries. 

\begin{figure}[h!]
\centering
\includegraphics[width=\linewidth]{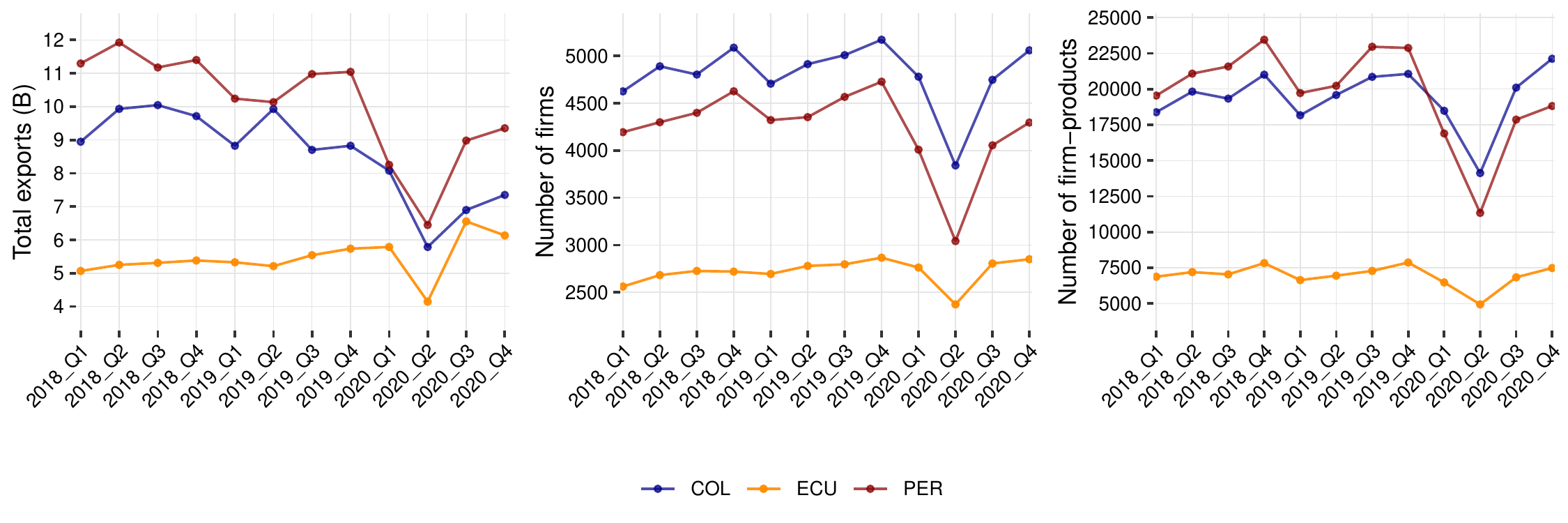}
\caption{Evolution of exports across countries. Left panel: total exports in billions. Middle panel: number of exporting firms. Right panel: number of firm-products.}
\label{fig:Aggregates fig1}
\end{figure}

Figure~\ref{fig:Aggregates fig1} shows the evolution of total exports, the number of exporting firms, and the total number of firm-product computed quarterly. For all countries, we clearly observe that the pandemic has led to a sharp decline in the second quarter of 2020. Comparing the second quarter of 2019 with the second quarter of 2020, we observe that total exports fell 42\% in Colombia, 20\% in Ecuador, and 36\% in Peru; the number of exporting firms fell 22\%, 15\%, and 30\%; and the number of firm-product dyads fell 28\%, 29\%, and 44\%, in Colombia, Ecuador, and Peru, respectively. The two latter measures recovered remarkably in the third and fourth quarters of 2020, bouncing back to similar levels observed in previous years. In contrast, the COVID--19 pandemic has left a negative impact on trade flows, which has been mainly reflected in the intensive margin of trade. In the Appendix, we add some evidence that the extensive margin effects were suffered mainly by smaller firms; see Table~\ref{tb:app_summary_size}.

For all countries, the distribution of the number of exported products at the firm level is stable and right-skewed. For example, for Colombia in 2019, 40.6\% of the companies are single-product, 18.6\% exported two products, 9.5\% three products, and 6.2\% four products. In Table~\ref{tb:summary}, we see that these values are similar for other years. A similar pattern is observed for Ecuador and Peru. 
\begin{table}[h!] 
\begin{center}
\caption{Summary statistics by levels of firm diversification.}
\label{tb:summary}\vspace{3pt}
\resizebox{\textwidth}{!}{\renewcommand{\arraystretch}{1.1}
\begin{tabular}{c cccccc cccccc cccccc}
\toprule
& \multicolumn{6}{c}{Colombia} & \multicolumn{6}{c}{Ecuador} & \multicolumn{6}{c}{Peru} \\
\cmidrule(lr){2-7}
\cmidrule(lr){8-13}
\cmidrule(lr){14-19}
Level & 1 & 2 & 3 & 4 & 5-10 & $>$10 & 1 & 2 & 3 & 4 & 5-10 & $>$10 & 1 & 2 & 3 & 4 & 5-10 & $>$10 \\
\midrule
Year & \multicolumn{18}{c}{Number of Exporting Firms} \\
\cmidrule(lr){1-1}  \cmidrule(lr){2-19}
2018 & 3,123 & 1,383 & 733 & 495 & 1,053 & 795 & 2,106 & 608 & 300 & 187 & 373 & 292 & 2,627 & 1,123 & 676 & 415 & 1,151 & 1,091 \\
2019 & 3,110 & 1,421 & 725 & 477 & 1,105 & 814 & 2,244 & 662 & 288 & 183 & 377 & 290 & 2,836 & 1,167 & 635 & 478 & 1,156 & 1,072 \\
2020 & 3,060 & 1,351 & 699 & 427 & 1,064 & 788 & 2,227 & 651 & 284 & 173 & 403 & 229 & 2,675 & 1,047 & 545 & 405 & 1,010 & 841 \\
\cmidrule(lr){2-19}
 & \multicolumn{18}{c}{Share of Exporting Firms (\%)} \\
\cmidrule(lr){2-19}
2018 & 41.2 & 18.2 & 9.7 & 6.5 & 13.9 & 10.5 & 54.5 & 15.7 & 7.8 & 4.8 & 9.6 & 7.6 & 37.1 & 15.9 & 9.5 & 5.9 & 16.3 & 15.4 \\
2019 & 40.6 & 18.6 & 9.5 & 6.2 & 14.4 & 10.6 & 55.5 & 16.4 & 7.1 & 4.5 & 9.3 & 7.2 & 38.6 & 15.9 & 8.6 & 6.5 & 15.7 & 14.6 \\
2020 & 41.4 & 18.3 & 9.5 & 5.8 & 14.4 & 10.7 & 56.1 & 16.4 & 7.2 & 4.4 & 10.2 & 5.8 & 41.0 & 16.1 & 8.4 & 6.2 & 15.5 & 12.9 \\
\cmidrule(lr){2-19}
& \multicolumn{18}{c}{Share of total exports (\%)} \\
\cmidrule(lr){2-19}
2018 & 27.9 & 12.7 & 31.6 & 3.9 & 13.2 & 10.7 & 27.0 & 42.4 & 9.4 & 4.0 & 9.3 & 7.9 & 14.9 & 7.9 & 8.2 & 5.4 & 38.8 & 24.8 \\
2019 & 24.5 & 12.9 & 2.5 & 8.6 & 34.8 & 16.7 & 27.5 & 9.3 & 12.6 & 34.4 & 11.3 & 4.8 & 14.6 & 12.5 & 3.3 & 12.8 & 32.5 & 24.3 \\
2020 & 22.8 & 14.9 & 7.0 & 4.2 & 31.3 & 19.7 & 34.6 & 31.1 & 6.9 & 12.1 & 8.8 & 6.7 & 16.6 & 9.9 & 6.5 & 7.6 & 41.5 & 18.0 \\
\cmidrule(lr){2-19}
& \multicolumn{18}{c}{Firms average number of destinations} \\
\cmidrule(lr){2-19}
2018 & 1.9 & 2.9 & 3.5 & 3.5 & 4.6 & 7.1 & 5.7 & 5.3 & 5.7 & 6.1 & 5.1 & 5.8 & 1.7 & 2.4 & 3.0 & 3.1 & 4.2 & 5.7 \\
2019 & 1.9 & 2.9 & 3.2 & 3.8 & 4.7 & 6.9 & 5.6 & 5.9 & 5.9 & 5.9 & 5.1 & 5.6 & 1.8 & 2.4 & 2.8 & 3.3 & 4.1 & 5.9 \\
2020 & 1.8 & 3.0 & 3.1 & 3.9 & 4.7 & 6.8 & 5.6 & 7.1 & 5.5 & 5.6 & 5.4 & 5.3 & 1.8 & 2.4 & 3.2 & 3.5 & 4.3 & 5.7 \\
\cmidrule(lr){2-19}
& \multicolumn{18}{c}{Firms total exports median (USD thousands)} \\
\cmidrule(lr){2-19}
2018 & 25.0 & 61.8 & 92.4 & 92.9 & 165.6 & 449.3 & 96.1 & 143.2 & 236.2 & 388.6 & 209.7 & 153.4 & 40.3 & 72.7 & 88.1 & 96.9 & 141.8 & 267.8 \\
2019 & 25.7 & 65.4 & 98.0 & 109.6 & 163.8 & 453.5 & 100.4 & 153.9 & 197.6 & 279.9 & 238.2 & 133.2 & 41.8 & 59.9 & 79.0 & 117.6 & 134.5 & 325.0 \\
2020 & 25.7 & 66.3 & 83.1 & 140.5 & 184.9 & 442.3 & 124.9 & 266.8 & 206.6 & 221.2 & 239.5 & 169.4 & 55.1 & 79.9 & 98.8 & 112.9 & 145.8 & 255.9 \\
\bottomrule
\end{tabular}}
\end{center}
\end{table}

Single-product companies account for a significant share of total exports: 24.5\% in Colombia, 27.5\% in Ecuador, and 14.6\% for Peru in 2019. However, they represent much larger shares of total exporting firms (40.6\%, 55.5\%, and 38.6\% for Colombia, Ecuador, and Peru, respectively). This evidence confirms that multi-product firms account for the bulk of trade flows for our set of developing countries. Notably, firms with more than four products account for 55.5\%, 45.7\%, and 56.8\% of total exports in Colombia, Ecuador, and Peru, respectively. Furthermore, single-product companies export on average to fewer destination markets than multi-product firms. For example, on average, for Colombia, single-product firms export to two destination markets, while those firms that export more than ten products do so to 7 markets. A similar pattern is observed in Peru. It is noteworthy that, for Ecuador, the number of export destinations is very similar for firms with different levels of diversification.

The evidence presented thus far confirms the main features highlighted in the literature: trade flows are indeed dominated by a minority of large firms that are highly diversified in terms of both products and destination markets. Furthermore, firm size is also positively related to product diversification. In Table~\ref{tb:summary}, we report the median of total exports for the sub-samples of firms with different diversification levels, showing that higher diversification implies higher export levels.\footnote{Here, we use the subsamples' median since high right skewness prevails even when dividing the population by the firms' diversification.}

\begin{figure}[h!]
\centering
\includegraphics[width=\linewidth]{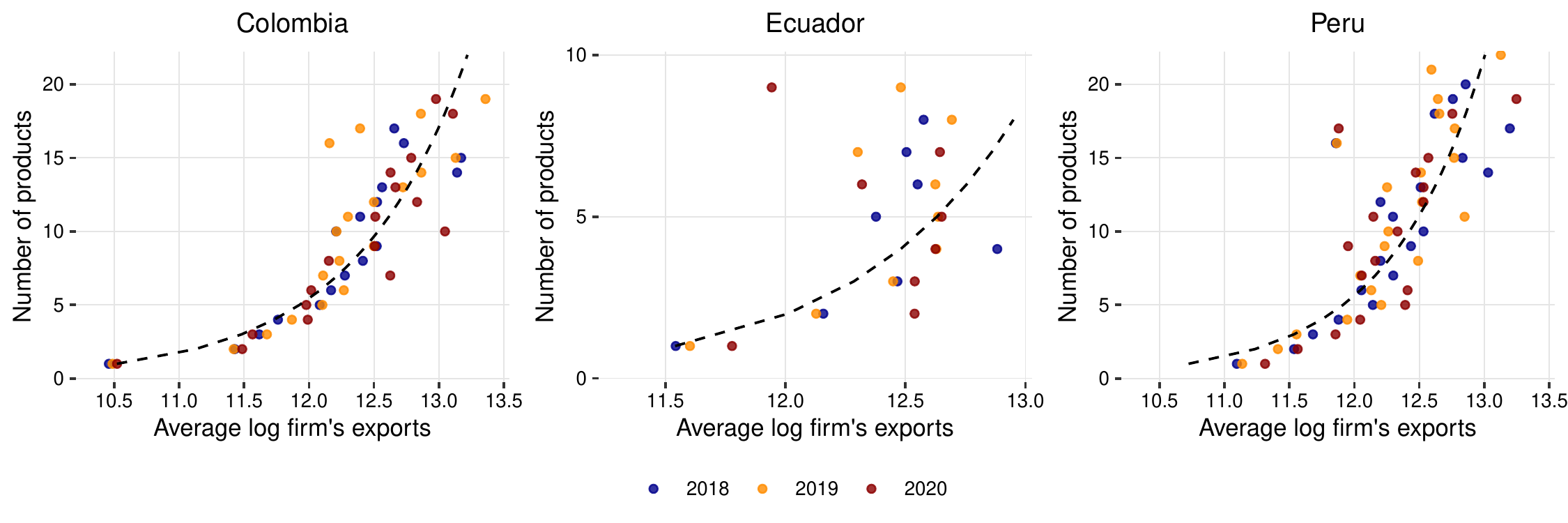}
\caption{Firms' diversification versus firms' size. Dots come from binned statistics. Observations are divided into bins, and then, for each bin, the average number of exported products and the average of the total exports in logs (which we use as a firm's size proxy) are computed. Dashed lines correspond to the exponential fitting of the binned results.}
\label{fig:size_np}
\end{figure}

In Figure~\ref{fig:size_np}, we further explore the relationship between firm size and the level of diversification. We find that, on average, the number of products grows exponentially with the firms' size (the natural logarithm of total exports). This evidence corroborates the contention of \cite{bottazzi2006gibrat} that a stochastic branching process can model the diversification behaviour of firms and that the number of products scales exponentially with the size of the firm. This is relevant to our study because it aligns with the capabilities literature discussed in Section~\ref{literature}, that firms diversify by branching out to products that require similar capabilities to what a firm already possesses. Therefore, this is \textit{prima facie} evidence confirming our conjecture about the importance of how closely a new product fits with the firm's export portfolio.

Before connecting more formally the evidence above, it is important to shed some light on the distribution of exports across product complexity levels, which we source from the Observatory of Economic Complexity.\footnote{We use data from 2015 to 2017 reported in \url{https://oec.world/en/rankings/pci/hs6/} and accessed: May 3, 2022. A shortcoming of the product complexity data is the missing values. Therefore, we first checked that there were no dramatic changes yearly, and then we averaged product complexity to consider as many products as possible.} We are interested in exploring whether the relationship between product complexity and export volume changes between products that are part of the core of firms' capabilities. A first way to study this is to rely on the concept of typical product vector (TPV), put forward by \cite{fontagne2018exporters}. In their approach, a firm's TPV is the vector of products (i.e., the product mix) that is the most commonly exported across the firms' destinations.\footnote{We refer the interested reader to \cite{fontagne2018exporters} for a full discussion. However, it is important to highlight that given that this vector of products is exported to at least two destinations, only firms that export to more than one country have a TPV. So, for the firms not falling within this category--a majority: 67\%--TPV is listed as missing.} 

Figure~\ref{fig:Complexity TPV} plots the distribution of export volumes across complexity levels. It shows that the distribution of firm-product exports changes only slightly across complexity quartiles. Interestingly, we find that products that are part of the firm's TPV tend to be exported in larger quantities across product complexity levels. This brings further support to the idea that product hierarchies exist and that not all products are the same within a firm's export portfolio. In \cite{fontagne2018exporters} the TPV should capture whether a firm is part of a bundle of products that are likely to be exported together and be part of the core capabilities of the firm. The TPV is a useful, data-driven measure, but it is not firmly grounded in a theory of firms' capabilities. As such, it offers no insight into how products within firms' portfolios relate to each other. Furthermore, its dichotomous nature does not allow measuring ``how'' core a product is with respect to a firm's capabilities. In the next section, we discuss these ideas more at length and put forward a novel measure of the coreness of products within firms' portfolios.
\begin{figure}[h!]
\centering
\includegraphics[width=\linewidth]{complexity_TPV_exports.pdf}
\caption{Violin plot of firm-product exports vs. product complexity, distinguishing between TPV and non-TPV products, for 2019.}
\label{fig:Complexity TPV}
\end{figure}

\subsection{Firm exports vector changes}
\label{subsec:changes}
We start by establishing the stability of the firms' export basket composition. If their composition is fickle, then there is little interest in understanding how it relates to the trade margin. But if, in contrast, export baskets are unlikely to change in the short term, then the extent to which products fit within them can be considered a relevant factor for exporting. Our descriptive evidence in Figure~\ref{fig:Aggregates fig1} suggests that the COVID--19 crisis mostly affected firms on their intensive, rather than extensive margin, lending support to the latter conjecture.

To measure the inter-temporal changes in firms' export baskets, we use the Bray \& Curtis similarity index, which accounts for the changes in the concentration of the export vector.\footnote{This measure has been previously used in the literature \cite[see:][for an application of the Bray \& Curtis similarity index]{fontagne2018exporters}.} Thus, let $X_i^t \in \mathbb{R}^K$ be the vector of exported products by firm $i$ in time $t$, where $K$ equals the total number of products from the HS codes at 6-digits. The Bray \& Curtis similarity is defined as:
\begin{equation}
\label{eq:BC}
BC_{it} = 1 - \dfrac{\sum_{k} |x_{ik}^t-x_{ik}^{t-1}|}
{\sum_{k}(x_{ik}^t+x_{ik}^{t-1})} ,
\end{equation}
where $x_{ik}^t=X_{ik}^t/\sum_{k'} X_{ik'}^t$ is the product's $k$ export share in firm $i$ at time $t$. The index is defined in the interval $[0,1]$. It will take values close to one when there are little changes in the export basket concentration profile and values closer to zero otherwise. 

\begin{figure}[h!]
\centering
\includegraphics[width=\linewidth]{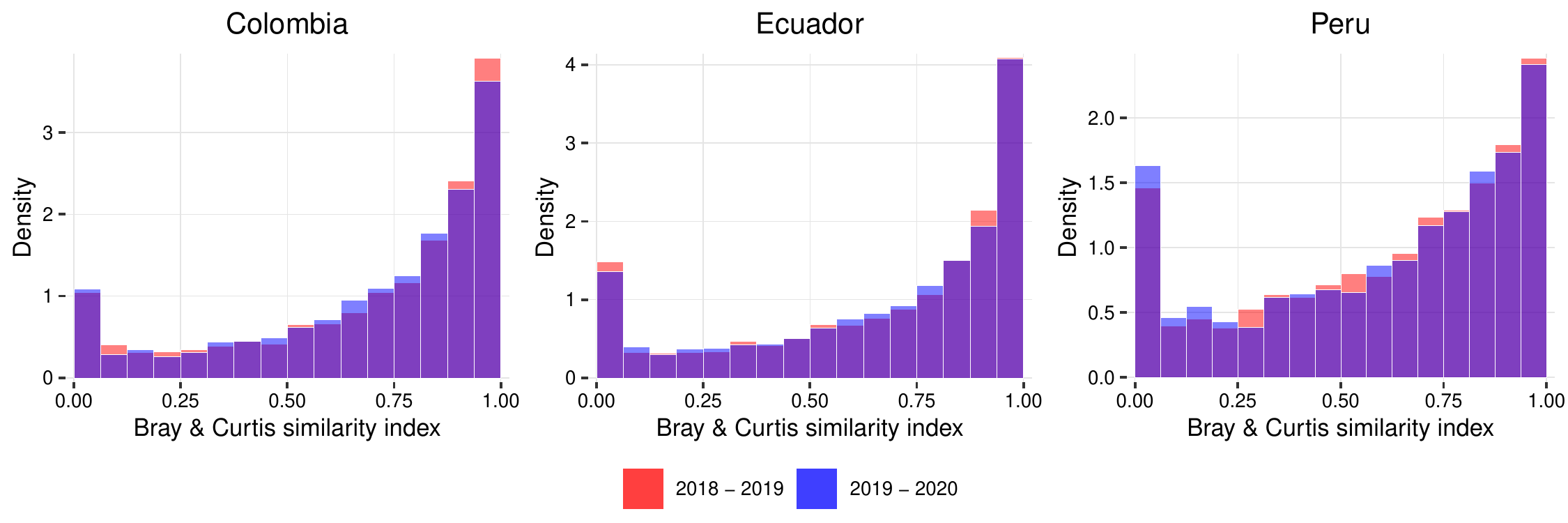}
\caption{Changes in firms export baskets between $t$ and $t-1$, histograms of the Bray \& Curtis similarity index.}
\label{fig:BC_hist}
\end{figure}

We analyse changes in the production vector using the Bray \& Curtis index for the periods 2018-2019 and 2019-2020, keeping in mind that the effects of the COVID--19 pandemic would be observed for this latter period. Figure~\ref{fig:BC_hist} overlays the distribution of changes in firms' export vectors for both periods. For a large part of the firms' population, there is a high similarity of the export baskets in consecutive years. However, the distribution is left-skewed with an excess probability mass towards zero (hinting at orthogonal changes of the exports vector). This means that while most firms do not change export vectors, there is a non-negligible share of firms that dramatically change the export basket. We found that most firms exhibiting such extreme changes ($BC \rightarrow 0$) have either low diversification or exited the export market.

Therefore, we find that firms tend to preserve the composition of their export baskets, even amid the COVID--19 crisis, offering more formalised evidence in support of \cite{benguria2021collpase} and the descriptive evidence in Figure~\ref{fig:Aggregates fig1}. Given that firms are unlikely to radically change their export portfolios in the short run, it becomes important to asses how each product fits within this and how this is related to firms' export performance. 

\subsection{Coreness}
\label{coreness_measure}

We have discussed in Section~\ref{literature} that an identified stylised fact in the literature is that typically firms have well-distinguished best-seller products that account for most of their exports \citep[see, e.g., ][]{eckel_neary2010}. Furthermore, we have also seen that diversification is positively related to the firm's size \citep{bottazzi2006gibrat}, highlighting the importance of core capabilities in the diversification process. 

Following the capability-based literature, we start from the premise that firms specialise in bundles of products that share similar capability requirements and that many firms export simultaneously. Therefore, in defining our product coreness measure, it is first necessary to determine the degree of proximity between products. Second, this proximity measure is used to assess how embedded the product is within firm's export bundles -- as we discuss in more details further below.

As our first step to obtain a measure of proximity among products, we ground our analysis on the bipartite export matrix. Thus, let $\mathbf{X}^t$ be the bipartite export matrix with dimensions $F\times K$, where rows represent the $F$ firms, columns are the $K$ products, and non-zero entries $X_{ik}^t$ are the exports of firm $i$ of product $k$ in year $t$. Then, we determine the relevant exported products for each firm, i.e., firms' product specialisation. To do this, we use the Revealed Comparative Advantage (RCA),\footnote{Identifying the relevant product eliminates around 5\% of the entries of the bipartite matrix because firms are generally very specialised in what they export. This is essential since considering any export can lead to irrelevant product-product relationships.} 
\begin{equation}
    RCA_{ik} = \frac{X_{ik}/\sum_{k'} X_{ik'}}{\sum_{i'} X_{i'k}/\sum_{i'}\sum_{k'} X_{i'k'}}.
\end{equation} 
We obtain the RCA-filtered bipartite matrix $Y$ whose generic entry $y_{ik}$ reads:
\begin{equation}
\label{eq:Y}
y_{ik}=
  \begin{cases}
  0 \text{ if } RCA_{ik}<1, \\ 
  1 \text{ if } RCA_{ik}\geq 1.
  \end{cases}
\end{equation}

We use the Jaccard index to measure the proximity between products.\footnote{The Jaccard index has been widely used as a relatedness measure to detect co-occurrences in data sets \citep[see:][]{leydesdorff2008normalization, boschma2014scientific, campi2021specialization}. \cite{hidalgo2007} use another alternative to measure relatedness between products. Our results are robust to both measures and are available upon request.} Thus, the proximity $J_{kk'}^t$ between products $k$ and $k'$ in year $t$ reads: 
\begin{equation}
\label{eq:jacc1}
J_{kk'}^t=\dfrac{\Lambda_{kk'}^t}{\Lambda_k^t+\Lambda_{k'}^t-\Lambda_{kk'}^t},
\end{equation}
where $\Lambda_{kk'}^t=\sum_i y_{ik}^ty_{ik'}^t$ is the number of times two different firms are relevant exporters of products $k$ and $k'$ together, and $\Lambda_{k}^t=\sum_i y_{ik}^t$ is the total number of firms that are relevant exporters of product $k$. $J_{kk'}^t$ takes values in the interval $[0,1]$, and it is defined as $J_{kk}=1$ for any $k$. 

High proximity for a product pair reveals they share similar capability requirements for their production. Mathematically, this happens when two products are exported together frequently by a group of firms. The literature argues that this high co-occurrence is a proxy that the firms have similar capabilities, which also implies that these products require a similar set of capabilities embedded in the firms. In other words, the co-occurrence reveals capabilities that are impossible to observe but are reflected in the joint production and consequent export.

The matrix $\mathbf{J}^t$ represents the product-product proximity network, where nodes are products and weighted links $J_{kk'}^t$ measure the similarity between them (we build this matrix at the firm level of each country independently). 

Finally, we use the matrix $\mathbf{J}^t$ to build our novel \textit{coreness} indicator, which takes the average degree of proximity between a product and all the other products in the export basket at the firm level, weighted on export flows. The coreness of a given product within a firm's export basket is defined as:
\begin{equation}
\label{eq:coreness}
Coreness_{ik}^t= \dfrac{\sum_{k'} J_{kk'}^t X_{ik}^t X_{ik'}^t }{\sum_{k'} X_{ik}^t X_{ik'}^t }.
\end{equation}
It takes values in the interval $[0,1]$, with higher values indicating that a product is embedded in a basket of highly related products, considering export flows as weights. 

\begin{figure}[h!]
\centering
\includegraphics[width=\linewidth]{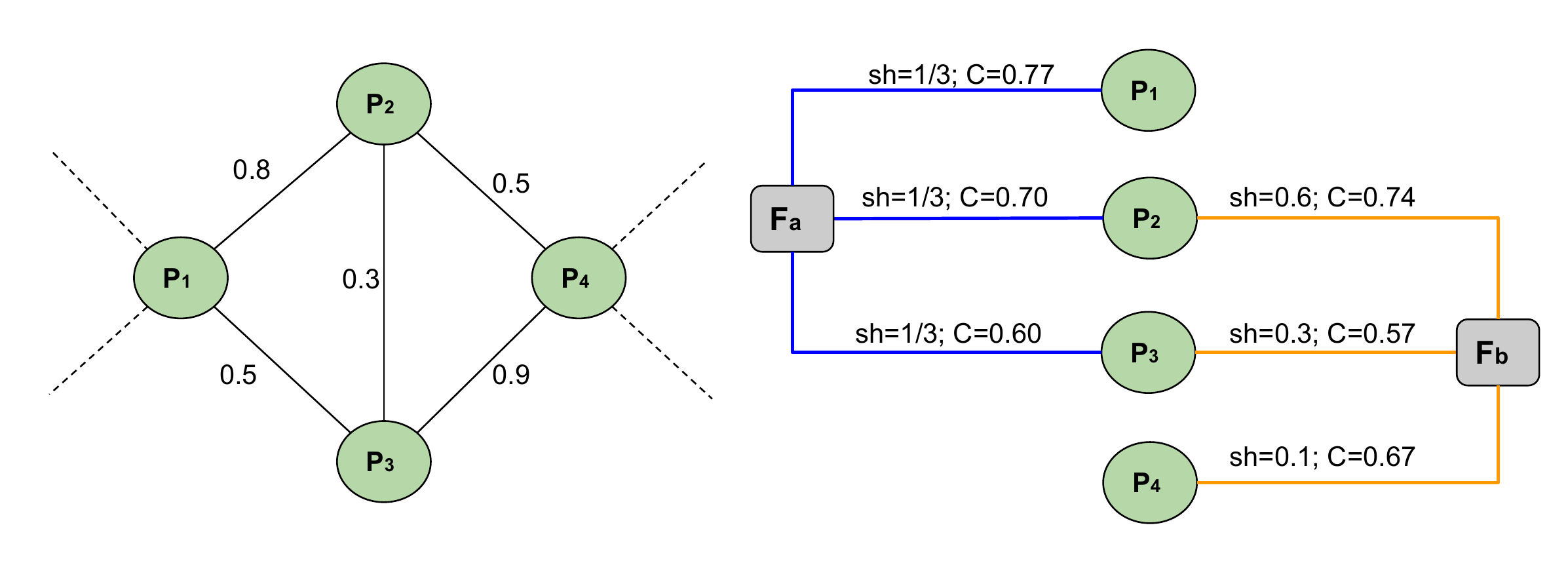}
\caption{Deriving coreness: a schematic example. Left panel shows a fraction of the products proximity network, where nodes represent products and link weights the proximity value. Right panel shows the product baskets of firms Fa and Fb, where ``sh'' is the export share and ``C'' is the corresponding computed coreness, see Eq.\eqref{eq:coreness}.}
\label{fig:example}
\end{figure}

In Figure~\ref{fig:example}, we show a schematic example of the product coreness calculation. In the left panel, we represent a portion of the product space or products proximity network (which we derive for the whole population using the Jaccard index). In this example, the most similar products are $P_3$ and $P_4$, $J_{34}=0.9$, and the less similar are $P_1$ and $P_4$, $J_{14}=0$. In the right panel there are two firms, Fa and Fb, which export three products each. The structure of the firms' baskets is different; while Fa uniformly distributes its exports  (sh = 1/3), Fb reveals a clear bias towards exporting $P_2$. For Fa the product $P1$ has the highest coreness (0.77). Note that even though the distribution of exports across products for Fa is uniform, $P1$ has higher coreness because -- based on the product space computed on the firms' population -- it is the product with the highest proximity to the other products exported by Fa. On the contrary, for Fb, $P2$ is the product with the highest coreness (0.74) and largest export shares. However, the coreness for $P3$ is less than for $P4$, despite the former accounting for a larger share of exports than the latter. This is because $P4$ is more strongly connected to $P2$ in the product space than $P3$.

\section{Results}
\label{results}

In this section, we put forward some key descriptive statistics on how our novel measure of coreness relates to the extensive and intensive margins of trade as well as its interaction with economic complexity. We also explore, through an econometric exercise, how coreness and export volumes are related, both at the firm-product and country-product levels.

\subsection{Coreness, export diversification and complexity}
\label{coreness_desc}
\begin{figure}[h!]
\centering
\includegraphics[width=\linewidth]{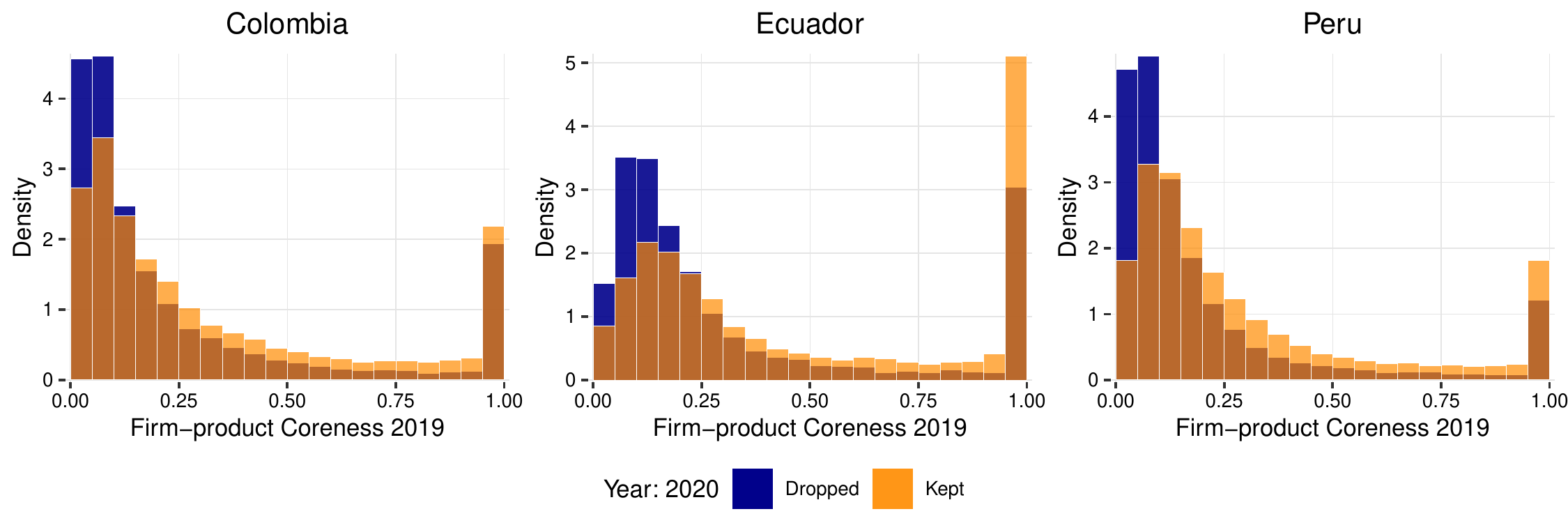}
\caption{Distribution of the product coreness in 2019 for products that are kept or dropped in 2020.}
\label{fig:Coreness_hist}
\end{figure}

Figure~\ref{fig:Coreness_hist} reports the distribution of coreness across firm-product in 2019 across the three countries. The figure overlays the distributions of products that have been dropped and those that have been kept in 2020. We want to see whether the population of persistent products in 2020 reveals differences with respect to the population of dropped products, in terms of the coreness observed in 2019. The results suggest that products at the core of firms' capabilities are less likely to be dropped. In 2019, the coreness density distribution of dropped products is more right-skewed than that of kept products. In other words, lower coreness values are typically more related to products that exit the market. We test this conjecture more formally within an econometric model in Table~\ref{tb:logit} in the Appendix.

\begin{figure}[h!]
\centering
\includegraphics[width=\linewidth]{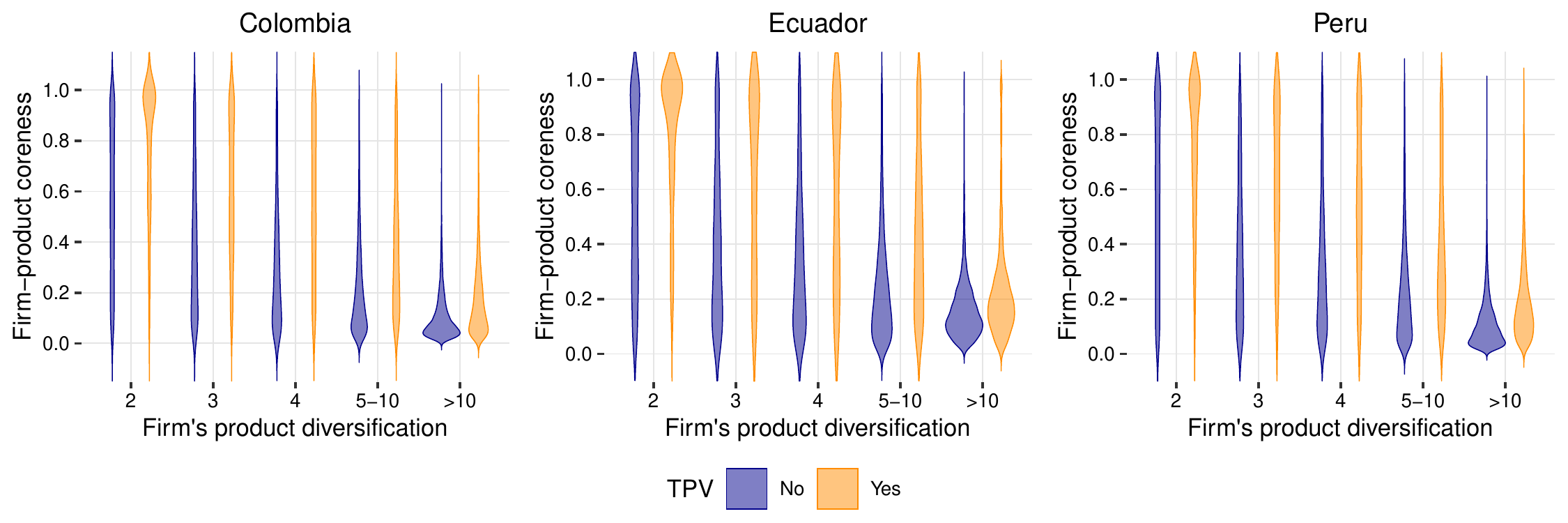}
\caption{Violin plot of coreness vs. firm diversification, distinguishing between TPV and non-TPV products, for 2019.}
\label{fig:Coreness_diversification}
\end{figure}

In light of the evidence above, we explore how coreness is distributed across firms with different levels of diversification. Figure \ref{fig:Coreness_diversification} reports the distribution of coreness by firms' product diversification, distinguishing between TPV and non-TPV products. We find striking differences in the distribution of coreness both along different levels of diversification and across TPV and non-TPV products. In particular, as the firms' number of exported products increases, the distribution of coreness is increasingly skewed towards lower levels. This implies that product coreness among highly diversified firms tends, on average, to be lower. It is also interesting to note that this is less the case for TPV products than for non-TPV ones. Products with higher coreness also tend, on average, to be part of the bundle of products that are most commonly exported across destinations, and are, therefore, part of the TPV. This further corroborates our contention that our measure of coreness captures the distance of products from firms' core capabilities.

\begin{figure}[h!]
\centering
\includegraphics[width=\linewidth]{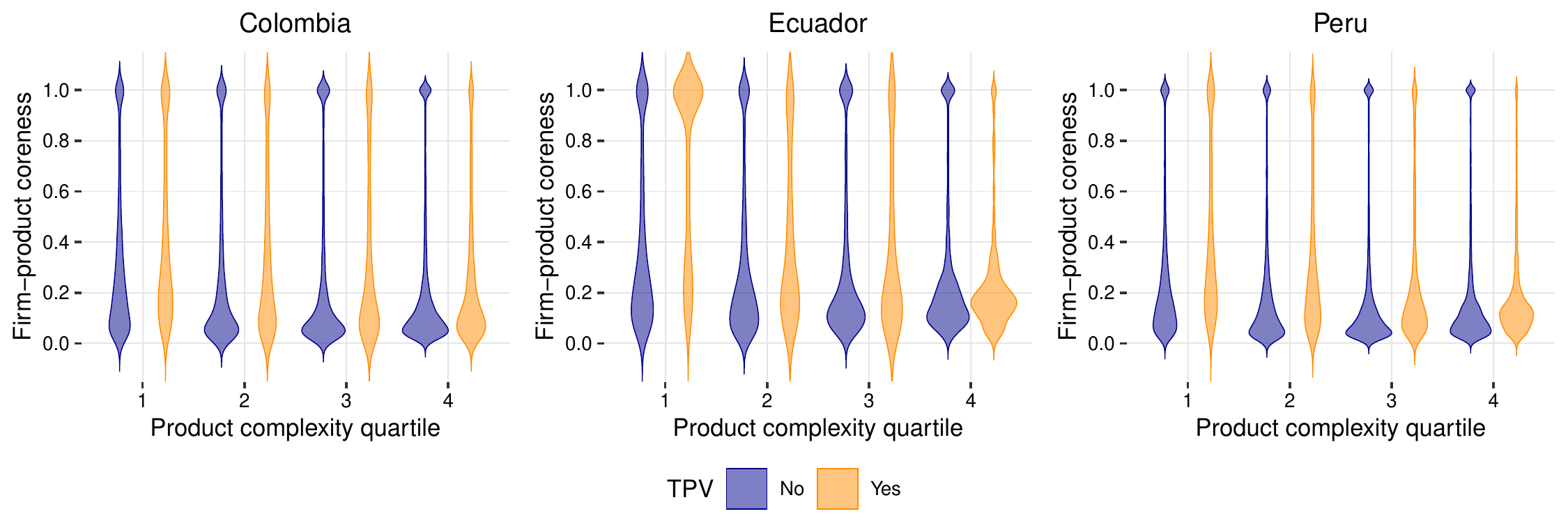}
\caption{Violin plot of coreness vs. product complexity, distinguishing between TPV and non-TPV products, for 2019.}
\label{fig:Coreness_complexity}
\end{figure}

Finally, Figure~\ref{fig:Coreness_complexity} reports the distribution of coreness by different levels of product complexity, distinguishing between TPV and non-TPV products. Similarly to the results in Figure \ref{fig:Coreness_diversification}, we find that as product complexity increases, the distribution of coreness is more right-skewed. Also, we find that this is less the case for TPV products.\footnote{We also observe that in all distributions, there is a significant portion of firm-products with coreness 1. As already pointed out, these firms are very specialised in a few products and are present across all levels of product complexity.}

These results are to be expected as we know that complexity is tightly related to export diversification at the country level \citep{hausmann2011network}, and, therefore, it makes sense that highly diversified firms export complex products with lower coreness. It is also reassuring that, on average, we find again that products exported to many destinations (and thus belonging to the firms' TPV) have higher coreness.

The evidence above hints at an intricate relationship between coreness, diversification, complexity, and export volumes. Larger export volumes are associated with export diversification, which in turn is associated with higher complexity. Within this, coreness seems to be negatively correlated with diversification: as more products are added to a firm's mix, the coreness of products in the mix decreases. Thus, larger firms have a more diversified and complex export portfolio, and yet most of their products have lower coreness. It is in fact reasonable to expect that as a firm includes new products, this will lower the coreness of the products in its export mix, although this does not mean that coreness loses importance as firms diversify.

We contend in fact that, \textit{ceteris paribus}, products at the core of firms' capabilities will be exported in larger quantities. To explore this hypothesis while properly accounting for firm characteristics, diversification, and product complexity, we turn to our econometric analysis in the following section.

\subsection{Coreness and exports: regression analysis}
\label{econometric}
We aim to further investigate our conjecture about the importance of coreness for exports while also exploring the structural relationships with complexity and diversification. Our main aim here is to establish the existence of structural relationships between coreness and exports without making a causal investigation, which we leave for further research. We do so both at the firm-product and country-product levels.

Concerning the former, we estimate the following benchmark model:
\begin{equation}\small
\label{eq:model}
\begin{split}
    y_{ikt} = \:\exp( 
    & \beta_{0} + 
    \beta_{1} \: \ln y_{ik(t-1)} +
    \beta_{2} \: Coreness_{ik(t-1)} + 
    \beta_{3} \: Coreness_{ik(t-1)}\times C_{20} + \\ 
    & \beta_{4} \: Complexity_{k} +
    \beta_{5} \: TPV_{ik(t-1)} +
    \beta_{6} \: \ln ND_{i(t-1)} +  
    \beta_{7} \: \ln NP_{i(t-1)} + \\ 
    & \tau_{t} + \gamma_{ik_{hs}})\:\eta_{ikt}\;;
\end{split}
\end{equation}
where, $y_{ikt}$ is exports for firm $i$ and product $k$ in year $t$, and $y_{ik(t-1)}$ is the corresponding variable lagged. We consider the lag to account for the persistence of exports over time. We also include key variables of interest and some additional controls on the right-hand side. $Coreness$ is our measure of interest, capturing how close a product lies to the firm's capabilities, which we lag by a year and also include an additional interaction with the year dummy $C_{20}$. This allows us to explore whether the relationship between exports and coreness changes over time, especially during the COVID--19 shock. In other words, we are interested in determining if, due to the lockdown policies generated by the pandemic, firms reshuffled the composition of the export basket, for instance, giving less relevance to core products.

$Complexity$ is product complexity, which we retrieve from the Observatory of Economic Complexity. TPV is a dummy variable taking value 1 if product $k$ in year $t-1$ is part of the typical product vector of firm $i$. $NP$ is the firm's number of products exported, capturing firms' product diversification. $ND$ is the number of firm export destinations. This is a different kind of firm-level diversification but it is important to include it in our analysis because it is likely related to the TPV dummy and captures firms' size and exposure to demand-side shocks in destination markets. We also include a set of fixed effects to control for time $\tau_t$ and firm-product HS chapter $\gamma_{ik_{hs}}$.\footnote{We identify HS chapter at 2 digits of the 2017 HS classification. Complexity is instead defined at the full 6-digits of the HS Classification.} Finally, $\eta_{ikt}$ is the error term, whose mean conditional to explanatory variables obeys $\text{E} [\eta_{ikt}|\cdot ]=1$.

We use a Poisson Pseudo Maximum Likelihood (PPML) estimation method as standard in the empirical trade literature since trade flows often take values equal to zero. Indeed, a firm might stop exporting a given product. In the estimation sample, we take into account all the firm-product pairs that had non-zero exports in $t-1$ (i.e., $y_{ik(t-1)}>0$), and therefore in the case that the pair exits the market, we keep considering it as $y_{ikt}=0$.

To facilitate the reading and interpretation of the results, we test our model considering all countries in the same sample, augmenting the model with country dummies. Nonetheless, in Appendix Table~\ref{tb:app_ppml}, we provide results by the three countries independently. The results are robust for each of the samples.

In Table~\ref{tb:ppml_firm_product_all}, we present the results of our benchmark model (column 1), which we also replicate by dividing products by complexity quartile (columns 2-5). Performing our analysis across complexity quartiles allows investigating how relationships among our variables of interest change along the complexity spectrum in a more informative way. Our benchmark model only includes multi-product firms, as single-product firms all have coreness equal to unity. Nonetheless, as a robust check, we replicate our analysis by including single-product firms and augmenting the model with a dummy capturing this, which we report in the Appendix Table~\ref{tb:app_ppml_single}. 
\begin{table}[h!]
\begin{center}
\scriptsize 
\caption{Firm-product level analysis. Estimation results all countries pooled}
\label{tb:ppml_firm_product_all}\vspace{3pt}
\resizebox{0.8\textwidth}{!}{\renewcommand{\arraystretch}{1.2}
\begin{tabular}{l ccccc} 
\toprule
Model &	(1)	&	(2)	&	(3)	&	(4)	&	(5)	\\
\midrule
 & Full sample & \multicolumn{4}{c}{Complexity Quartile}  \\
 \cmidrule(lr){2-2}
\cmidrule(lr){3-6} 
 &  & 1st & 2nd & 3rd & 4rd \\
\midrule
Exports (log of) (t-1) & 0.568*** & 0.507*** & 0.463*** & 0.748*** & 0.555*** \\
 & (0.060) & (0.083) & (0.113) & (0.025) & (0.057) \\
Coreness (t-1) & 1.314*** & 1.527*** & 1.736*** & 0.610*** & 1.773*** \\
 & (0.276) & (0.393) & (0.545) & (0.163) & (0.419) \\
Coreness (t-1) $\times$ $C_{20}$ & 0.057 & 0.109 & 0.228 & 0.043 & -0.223 \\
 & (0.121) & (0.161) & (0.186) & (0.106) & (0.191) \\
Complexity & -0.135** & -0.161 & -0.481 & -0.117 & -0.077 \\
 & (0.067) & (0.107) & (0.386) & (0.122) & (0.156) \\
TPV (t-1) & 0.151** & 0.128* & 0.304* & -0.034 & 0.218** \\
 & (0.062) & (0.070) & (0.169) & (0.078) & (0.090) \\
Firm's number of destinations (log of) (t-1) & -0.137 & -0.128 & 0.029 & -0.057 & 0.012 \\
 & (0.129) & (0.148) & (0.182) & (0.128) & (0.184) \\
Firm's number of products (log of) (t-1) & -0.093* & -0.098* & 0.048 & 0.008 & -0.211 \\
 & (0.051) & (0.057) & (0.118) & (0.071) & (0.201) \\
$C_{20}$ & -0.286*** & -0.338** & -0.342** & -0.240*** & -0.193 \\
 & (0.100) & (0.140) & (0.151) & (0.060) & (0.147) \\
Constant & 7.438*** & 8.558*** & 7.294*** & 3.845*** & 6.513*** \\
 & (0.995) & (1.382) & (1.313) & (0.437) & (0.693) \\
Observations & 133,591 & 54,450 & 25,884 & 21,550 & 22,913 \\
Firm-HS2 FE & Yes & Yes & Yes & Yes & Yes \\
\bottomrule
\multicolumn{6}{p{14cm}}{\textit{Notes:} The dependent variable are exports for firm $i$ and product $k$ in year $t$. Robust standard errors in parentheses. Significance level: *** p$<$0.01, ** p$<$0.05, * p$<$0.10.}
\end{tabular}}
\end{center}
\end{table}

Coreness is positively related to export volumes overall and across complexity quartiles. The interaction with the 2020 dummy ($C_{20}$) is generally not statistically significant, suggesting that coreness' positive relationship with exports is unchanged during the COVID--19 pandemic. 

Complexity is negatively related to export volumes, suggesting that, overall, firms in our three countries export complex products in lower quantities than less complex ones. However, coefficients become insignificant when replicating our analysis by complexity quartile; this is expected as the source of variation in complexity our analysis can exploit is significantly curtailed when we look at each complexity quartile separately.

The TPV dummy also shows a consistently positive sign. This adds to, rather than dampens, the positive relationship between coreness and exports. The two measures capture two different features of how a product fits within a firm's export portfolio. Referring back to the theoretical discussion in section \ref{literature}, TPV captures those products that firms are most efficient at exporting and are therefore sold across multiple destinations. Coreness, in contrast, measures how closely a product lies with other products and, we argue, the firm's capabilities, irrespective of the number of destinations to which this is exported.

Product diversification does not have an overall significant coefficient. When it does, it is weakly significant and negative. This may be surprising, but we should bear in mind that we are looking at firm-product combinations, that we use firm dummies and that firms' export portfolios are stable over time. This means that although it is possible that as a firm includes more products in its portfolio, it increases its total export volume, however, each product will compete with the others, resulting in a decrease in export at the firm-product level.

Firms' diversification in terms of destination markets is also insignificant, as its effect at the firm-product level is likely absorbed by the TPV dummy and the lag of the outcome variable accounting for firms' fixed effects. 

\subsection{Country-level relevance of coreness}

The analysis thus far has shed light on how firm-product exports relate to coreness at the micro level, finding strong evidence that the two are positively correlated. This has, of course, key implications for firms, how they include new products in their portfolio and their export performance. We also wish to push our analysis further and draw some broader insights for the policymakers. The literature on economic complexity has highlighted that not all products are the same and that specialisation matters for countries' economic development. Taking stock of this, we replicate our empirical analysis at the product level across countries to investigate whether the coreness of firm-product combinations ``trickle up'', impacting country-level exports across products.

Therefore, we estimate the following linear model:
\begin{equation}\small
\label{eq:model_country}
\begin{split}
    \ln y_{ckt} = 
    & \: \beta_{0} + 
    \beta_{1} \: \ln y_{ck(t-1)} +
    \beta_{2} \: Coreness_{ck(t-1)} + 
    \beta_{3} \: Coreness_{ck(t-1)}\times C_{20} + \\
    & \beta_{4} \: Complexity_{k} +
    \beta_{5} \: TPV_{ck(t-1)} + 
    \beta_{6} \: \ln ND_{ck(t-1)} + 
    \beta_{7} \: \ln NP_{ck(t-1)} + \\
    & \beta_{8} \: SingleShare_{ck(t-1)} + 
    \tau_{t} + \gamma_{ck_{hs}} + \varepsilon_{ckt}.
\end{split}
\end{equation}
In this equation, each element has been averaged across firms across countries $c$ and product $k$, by using export values as weights. The outcome variable $y_{ckt}$ and its lag $y_{ck(t-1)}$ are product-level aggregations across firms exporting product $k$ in the country $c$.
The variables are unchanged, but they now capture the average firm-product characteristics under which each product is exported. $Coreness_{ck}$ as an average is computed only across multi-product firms, so that the mean is not skewed by single-product firms that have coreness equal to 1. However, to account for the importance that single-product firms represent in the export of each product $k$, we also include an additional control $SingleShare$. This is the share of exports of each product in each country that is provided for by single-product firms.

The variables for product and destination diversification ($NP$ and $ND$, respectively) are also averaged and therefore reflect the average firm-level diversification with which each product $k$ is exported. In the same way, $TPV$ becomes the share of export of each product $k$ that is exported while belonging to a firm's typical product vector.

In Table~\ref{tb:product_level}, we present our country-level results. Overall we find results in line with our firm-product level analysis. Products that are, on average, core to firms' capabilities are exported in larger quantities. Complexity remains negatively associated with export volumes, suggesting that our three countries struggle to export complex products. The share of product exports that is part of firms' TPV is, in contrast, positively associated with export flows, which is in line with our firm-level results. It is also noteworthy that the average number of destination markets ($ND$) of firms exporting a given product is associated with larger trade volumes, this is likely capturing the importance of firm's size for export volumes. Firms' average product diversification is not statistically significant when looking at all products. Furthermore, we find rather unstable coefficients when looking across complexity quartiles. The share of exports that single-product firms trade exhibits a positive coefficient: products exported by single-product firms are traded in larger volumes at the country level too; however, this result is not stable across complexity quartiles. These two results suggest that the relationship between aggregate export and firms' diversification changes with the products' complexity.
\begin{table}[ht!]
\begin{center}
\scriptsize 
\caption{Country-product level analysis. Estimation results all countries pooled}
\label{tb:product_level}\vspace{3pt}
\resizebox{0.8\textwidth}{!}{\renewcommand{\arraystretch}{1.2}
\begin{tabular}{l ccccc} 
\toprule
Model &	(1)	&	(2)	&	(3)	&	(4)	&	(5)	\\
\midrule
 & Full sample & \multicolumn{4}{c}{Complexity Quartile}  \\
 \cmidrule(lr){2-2}
\cmidrule(lr){3-6} 
 &  & 1st & 2nd & 3rd & 4rd \\
\midrule
Exports (log of) (t-1) & 0.742*** & 0.794*** & 0.745*** & 0.668*** & 0.675*** \\
 & (0.022) & (0.039) & (0.049) & (0.048) & (0.052) \\
Coreness (HS mean) (t-1) & 0.432** & 0.534 & 0.713* & -0.608 & 1.350*** \\
 & (0.199) & (0.408) & (0.376) & (0.426) & (0.475) \\
Coreness (HS6 mean) (t-1) $\times C_{20}$ & 0.151 & -0.117 & -0.077 & 1.122** & -0.013 \\
 & (0.215) & (0.332) & (0.436) & (0.436) & (0.540) \\
Complexity & -0.124** & -0.330** & -0.377 & -0.213 & -0.024 \\
 & (0.049) & (0.138) & (0.281) & (0.305) & (0.255) \\
TPV (HS6 mean) (t-1) & 0.440*** & 0.439** & 0.486** & 0.551** & 0.409* \\
 & (0.104) & (0.212) & (0.216) & (0.225) & (0.208) \\
Firm's number of destinations (log of HS6 mean) (t-1) & 0.164*** & 0.214** & 0.037 & 0.272** & 0.113 \\
 & (0.054) & (0.092) & (0.118) & (0.134) & (0.112) \\
Firm's number of products (log of HS6 mean) (t-1) & 0.031 & 0.017 & 0.181** & -0.198** & 0.199** \\
 & (0.042) & (0.090) & (0.090) & (0.082) & (0.098) \\
Single product firms' export share (HS6 mean) (t-1) & 0.955*** & 0.646* & 0.615 & 0.838* & 1.068 \\
 & (0.219) & (0.336) & (0.542) & (0.491) & (0.664) \\
$C_{20}$ & -0.239** & -0.134 & -0.354* & -0.497*** & -0.069 \\
 & (0.096) & (0.164) & (0.194) & (0.189) & (0.225) \\
Constant & 2.228*** & 1.260** & 1.776*** & 3.881*** & 2.083*** \\
 & (0.234) & (0.518) & (0.539) & (0.489) & (0.645) \\
Observations & 3,635 & 1,009 & 926 & 885 & 735 \\
R-squared & 0.769 & 0.845 & 0.769 & 0.743 & 0.653 \\
Country-HS2 FE & Yes & Yes & Yes & Yes & Yes \\
\bottomrule
\multicolumn{6}{p{15cm}}{\textit{Notes:} The dependent variable is total exports of product $k$ across countries $c$. All averages are computed at the product (HS6) and year level and weighted on export flows; robust standard errors in parentheses. Significance level: *** p$<$0.01, ** p$<$0.05, * p$<$0.10.}\end{tabular}}
\end{center}
\end{table}



\section{Concluding remarks}\label{conclusions}

We propose a novel measure of firm-product coreness that captures how embedded a product is within a firm's export portfolio. We show that products with a high coreness are less likely to be dropped over time and that they are traded in larger volumes. These firm-level patterns are also reflected in countries' export structures and our results are robust to periods of crisis, such as those caused by the COVID--19 outbreak in 2020. 


Our measure of coreness is grounded in the literature on firms' capabilities \citep{penrose1959,teece1994} and the importance of its coherence \citep{brusoni2001,dosi2017firmsknow,dosi2022coherence}. In line with these theoretical insights, our results expand on the established fact that firms do not diversify randomly: by incorporating information on the product-product relatedness with our measure of coreness, we provide novel key elements to understand firm diversification in international trade. Beyond shedding new light on the extensive margin, we show that the coreness of a product within a company's export portfolio can explain variations in the intensive margin of trade, too. 

Our findings are relevant for understanding the relationship between companies' performance and the structure of their product portfolios. 
Our measure of coreness is in line with the idea of product basket coherence that has emerged in the recent literature: if a company exports several products and all of them have high coreness, then the company's export basket will have high coherence too.

However, our contribution goes further: we show that already at the firm-product level -- regardless of the overall coherence of a company's export portfolio -- the coreness of a product within a firm's portfolio plays a role in the company's performance. 

Finally, we show that these firm-level results also have implications at the aggregate level. Products that have on average high coreness within the firms that export them are also traded in larger quantities at the country level. This means that firm-product relationships have bearing on countries' specialisation. These results are particularly relevant for policies aiming at fostering specialisation trajectories towards high-tech, complex industries. Notably, if such attempts are performed by a few firms who make the leap towards new products that are different from what they already produce, they are unlikely to be successful and, as a result, shifts in countries' export mix won't occur. 

Two main policy recommendations emerge from our analysis. On the one hand, support should be given to firms in order to identify the products that they are most likely to successfully include, i.e. those with higher coreness within firms' existing export baskets. Second, governments should provide firms with support to expand their capabilities so that they can diversify and successfully include new hi-tech and complex products.

\clearpage
\newpage
\bibliographystyle{chicago}
\bibliography{biblio}

\clearpage
\newpage

\section*{Appendix}\label{sec:App}
\setcounter{table}{0}\renewcommand{\thetable}{A.\arabic{table}}
\setcounter{figure}{0}\renewcommand{\thefigure}{A.\arabic{figure}}

\subsection*{Summary statistics}
Table~\ref{tb:app_summary_size} presents the summary statistics. To highlight the heterogeneity in the evolution of exports, we have aggregated the results by groups of firms of similar sizes. Sizes are determined according to the quartiles of the distribution of total annual exports (in logs) at the firm level for each country independently. A common pattern is that the higher the size, the firms export more products to many different destinations.
\begin{table}[h!]\scriptsize 
\begin{center}
\caption{Summary statistics by firms size}
\label{tb:app_summary_size}\vspace{3pt}
\resizebox{0.8\textwidth}{!}{\renewcommand{\arraystretch}{1.05}
\begin{tabular}{c cccc cccc cccc}
\toprule
& \multicolumn{4}{c}{Colombia} & \multicolumn{4}{c}{Peru} & \multicolumn{4}{c}{Ecuador} \\
\cmidrule(lr){2-5}
\cmidrule(lr){6-9}
\cmidrule(lr){10-13}
Firm Size & S1 & S2 & S3 & S4 & S1 & S2 & S3 & S4 & S1 & S2 & S3 & S4 \\
\midrule
year & \multicolumn{12}{c}{Total exports (USD millions)} \\
\cmidrule(lr){1-1}
\cmidrule(lr){2-13}
2018 & 7.2 & 39.2 & 194.1 & 38,396.9 & 5.8 & 43.6 & 278.3 & 45,465.3 & 1.5 & 16.9 & 192.7 & 20,808.8 \\
2019 & 5.8 & 33.5 & 187.3 & 36,051.6 & 5.4 & 41.4 & 263.9 & 42,084.5 & 1.5 & 16.7 & 177.0 & 21,690.7 \\
2020 & 6.0 & 35.1 & 189.3 & 27,881.7 & 4.3 & 32.4 & 243.7 & 32,753.0 & 1.1 & 18.3 & 221.9 & 22,434.7 \\
\cmidrule(lr){2-13}
& \multicolumn{12}{c}{Number of exporting firms} \\
\cmidrule(lr){2-13}
2018 & 1,289 & 1,660 & 2,158 & 2,633 & 1,222 & 1,602 & 2,001 & 2,407 & 713 & 834 & 1128 & 1476 \\
2019 & 1,106 & 1,677 & 2,310 & 2,738 & 1,193 & 1,651 & 2,118 & 2,550 & 672 & 880 & 1,204 & 1,585 \\
2020 & 1,075 & 1,558 & 2,175 & 2,710 & 901 & 1,302 & 1,936 & 2,479 & 492 & 853 & 1,245 & 1,578 \\
\cmidrule(lr){2-13}
& \multicolumn{12}{c}{Firms average exports (USD thousands)} \\
\cmidrule(lr){2-13}
2018 & 5.6 & 23.6 & 89.9 & 14,583.0 & 4.7 & 27.2 & 139.1 & 18,888.8 & 2.1 & 20.3 & 170.8 & 14,098.1 \\
2019 & 5.2 & 20.0 & 81.1 & 13,167.1 & 4.5 & 25.1 & 124.6 & 16,503.7 & 2.2 & 19.0 & 147.0 & 13,685.0 \\
2020 & 5.6 & 22.5 & 87.0 & 10,288.4 & 4.7 & 24.9 & 125.9 & 13,212.2 & 2.3 & 21.4 & 178.2 & 14,217.2 \\
\cmidrule(lr){2-13}
& \multicolumn{12}{c}{Firms average number of products} \\
\cmidrule(lr){2-13}
2018 & 1.9 & 2.8 & 4.2 & 8.8 & 2.8 & 4.5 & 7.8 & 10.0 & 1.8 & 3.3 & 4.5 & 4.3 \\
2019 & 1.9 & 2.8 & 4.1 & 8.6 & 2.4 & 3.9 & 7.4 & 9.9 & 1.6 & 3.3 & 4.1 & 4.1 \\
2020 & 1.9 & 2.8 & 4.2 & 8.7 & 2.7 & 3.5 & 6.2 & 8.0 & 1.8 & 3.0 & 3.6 & 3.7 \\
\cmidrule(lr){2-13}
& \multicolumn{12}{c}{Firms average number of destinations} \\
\cmidrule(lr){2-13}
2018 & 1.1 & 1.4 & 2.1 & 6.3 & 1.1 & 1.3 & 2.0 & 5.9 & 1.1 & 1.5 & 3.1 & 11.2 \\
2019 & 1.1 & 1.3 & 2.0 & 6.2 & 1.1 & 1.3 & 1.9 & 5.8 & 1.1 & 1.4 & 3.0 & 11.1 \\
2020 & 1.1 & 1.3 & 1.9 & 6.1 & 1.1 & 1.2 & 1.8 & 5.5 & 1.1 & 1.6 & 3.5 & 10.7 \\
\bottomrule
\multicolumn{13}{p{15cm}}{\textit{Notes:} Sizes are defined by the quantiles of distribution of the total year exports (in logs). The cut points in thousands of US dollars are, Colombia, $Q_1=11.2$, $Q_2=41.4$, and $Q_3=194.9$; Peru, $Q_1 = 10.9$, $Q_2=54.1$, and $Q_3= 311.9$; and, Ecuador, $Q_1= 5.9$, $Q_2=48.9$, and $Q_3=448.3$.}
\end{tabular}}
\end{center}
\end{table}

\subsection*{Extensive margin: logit estimations}
Our analysis in the main text focuses on the intensive margin of trade, which our descriptive analysis in Figure \ref{fig:Aggregates fig1} shows to have been impacted the most by the COVID--19 pandemic. We wish, however, to complement this with a study of the extensive margin of trade too. This is because in Figure \ref{fig:BC_hist}, we do find that products that have been dropped often have lower coreness. To further explore this finding, we model the probability of exporting at the firm-product level and estimate the following benchmark model:
\begin{equation}\small
\label{eq:logit}
\begin{split}
    \text{Pr}[y_{ikt}=1|\cdot] = 
    \Lambda(
    & \beta_{0} + 
    \beta_{1} \: \ln y_{i(t-1)} +
    \beta_{2} \: Coreness_{ik(t-1)} + 
    \beta_{3} \: Coreness_{ik(t-1)}\times C_{20} + \\ 
    & \beta_{4} \: Complexity_{k} +
    \beta_{5} \: TPV_{ik(t-1)} + 
     \beta_{7} \: SingleProduct_{it} +
    \tau_{t} + \gamma_{k_{hs}t})\;;
    \end{split}
\end{equation}
where $y_{ikt}$ is a dummy that indicates whether the firm $i$ exports the product $k$ in year $t$, and $\Lambda(\cdot)$ is the logistic cdf. 

The estimations are shown in Table~\ref{tb:logit}. The results confirm our claim that coreness enhances the probability that a firm keeps exporting a product: the higher the coreness, the higher the probability of continuing to export. In addition, we show, for the pooled countries estimation (models 4 and 5), that the effect of coreness is slightly reduced when we include the interaction with the dummy of the crisis period. The effect of Complexity is less clear, it is positive for Colombia, negative for Ecuador, and not significant for Peru. The sign of the TPV dummy is positive, in agreement with \cite{fontagne2018exporters} findings. The effect for single-product firms is significant and negative. As we saw in Table~\ref{tb:app_summary_size}, the smaller companies are typically much less diversified in products and destinations. Thus, all these results provide more evidence to what is discussed in subsection~\ref{subsec:changes}. Accordingly, the excess probability mass observed for orthogonal or dramatic changes in the export basket is due to the dynamics of smaller companies that are less likely to keep their products in the foreign market. 
\begin{table}[h!]
\begin{center}
\scriptsize 
\caption{Estimation results for Colombia, Ecuador and Peru, countries and countries pooled.}
\label{tb:logit}\vspace{3pt}
\resizebox{0.7\textwidth}{!}{\renewcommand{\arraystretch}{1.2}
\begin{tabular}{l ccccc} 
\toprule
Model &	(1)	&	(2)	&	(3)	&	(4)	&	(5)	\\
\midrule
 & COL & ECU & PER & \multicolumn{2}{c}{Countries pooled}  \\
\midrule
Exports (log of) (t-1) & 0.266*** & 0.200*** & 0.254*** & 0.244*** & 0.244*** \\
 & (0.004) & (0.006) & (0.004) & (0.003) & (0.003) \\
Coreness (t-1) & 2.171*** & 1.888*** & 2.411*** & 2.193*** & 2.199*** \\
 & (0.047) & (0.070) & (0.045) & (0.029) & (0.034) \\
Year 2020 & -0.078*** & -0.181*** & -0.335*** & -0.213*** & -0.211*** \\
 & (0.017) & (0.027) & (0.015) & (0.010) & (0.014) \\
Year 2020 * Coreness (t-1) &  &  &  &  & -0.011 \\
 &  &  &  &  & (0.035) \\
Complexity & 0.031** & -0.085*** & -0.010 & -0.008 & -0.008 \\
 & (0.016) & (0.024) & (0.014) & (0.009) & (0.009) \\
TPV (t-1) & 1.545*** & 1.396*** & 1.305*** & 1.423*** & 1.423*** \\
 & (0.021) & (0.038) & (0.020) & (0.013) & (0.013) \\
Single product firm (t-1) & -0.991*** & -0.763*** & -1.330*** & -1.078*** & -1.078*** \\
 & (0.045) & (0.064) & (0.047) & (0.029) & (0.029) \\
Constant & -3.920*** & -4.265*** & -3.970*** & -3.630*** & -3.631*** \\
 & (0.274) & (0.554) & (0.323) & (0.189) & (0.189) \\
Observations & 77,868 & 29,013 & 97,959 & 204,866 & 204,866 \\
 \bottomrule
\multicolumn{6}{p{12cm}}{\textit{Notes:} Robust standard errors in parentheses. All estimations include fixed effect dummies for product HS chapter (first two digits of the HS code) and countries (models 4 and 5). Significance level: *** p$<$0.01, ** p$<$0.05, * p$<$0.10.}
\end{tabular}}
\end{center}
\end{table}

\clearpage
\newpage
\subsection*{Econometric estimations by country}

\begin{table}[h!]
\begin{center}
\scriptsize 
\caption{Estimation results for Colombia, Peru and Ecuador}
\label{tb:app_ppml}\vspace{3pt}
\resizebox{0.7\textwidth}{!}{\renewcommand{\arraystretch}{1.1}
\begin{tabular}{l ccccc} 
\toprule
Model &	(1)	&	(2)	&	(3)	&	(4)	&	(5)	\\
\midrule
 & Full sample & \multicolumn{4}{c}{Complexity Quartile}  \\
 \cmidrule(lr){2-2}
\cmidrule(lr){3-6} 
 &  & 1st & 2nd & 3rd & 4rd \\
\midrule
 \multicolumn{6}{c}{\textit{Estimation results for Colombia}} \\
\midrule
Exports (log of) (t-1) & 0.752*** & 0.682*** & 0.753*** & 0.806*** & 0.614*** \\
 & (0.029) & (0.051) & (0.036) & (0.029) & (0.037) \\
Coreness (t-1) & 1.051*** & 1.379*** & 0.854*** & 0.527** & 1.725*** \\
 & (0.212) & (0.341) & (0.279) & (0.205) & (0.360) \\
Year 2020 & -0.231** & -0.462** & -0.125** & -0.234*** & 0.121 \\
 & (0.104) & (0.188) & (0.058) & (0.074) & (0.163) \\
Year 2020 * Coreness (t-1) & -0.034 & 0.164 & 0.013 & 0.096 & -0.553*** \\
 & (0.129) & (0.201) & (0.095) & (0.139) & (0.187) \\
Complexity & -0.017 & 0.100 & -0.063 & -0.049 & 0.230 \\
 & (0.054) & (0.105) & (0.146) & (0.141) & (0.214) \\
TPV (t-1) & -0.027 & -0.010 & -0.155 & -0.158 & 0.251** \\
 & (0.060) & (0.098) & (0.095) & (0.096) & (0.113) \\
Number of destinations (log of) (t-1) & 0.163 & 0.277 & -0.024 & -0.321* & -0.103 \\
 & (0.141) & (0.187) & (0.137) & (0.175) & (0.211) \\
Firm's number of products (log of) (t-1) & -0.137*** & -0.106* & -0.158 & 0.029 & -0.063 \\
 & (0.053) & (0.061) & (0.140) & (0.104) & (0.208) \\
Constant & 3.784*** & 4.935*** & 4.135*** & 3.738*** & 5.200*** \\
 & (0.630) & (1.033) & (0.621) & (0.596) & (0.961) \\
Observations & 53,263 & 18,903 & 10,707 & 10,290 & 10,116 \\
\midrule
 \multicolumn{6}{c}{\textit{Estimation results for Ecuador}} \\
\midrule
Exports (log of) (t-1) & 0.394*** & 0.376*** & 0.098 & 0.553*** & 0.499*** \\
 & (0.060) & (0.066) & (0.082) & (0.042) & (0.111) \\
Coreness (t-1) & 1.816*** & 2.540*** & 2.682*** & 0.405 & 0.763 \\
 & (0.351) & (0.429) & (0.962) & (0.343) & (0.697) \\
Year 2020 & -0.295 & -0.243* & -0.256 & -0.329* & -0.388* \\
 & (0.186) & (0.128) & (0.395) & (0.168) & (0.217) \\
Year 2020 * Coreness (t-1) & 0.124 & 0.049 & 0.540 & 0.003 & 0.251 \\
 & (0.223) & (0.155) & (0.479) & (0.231) & (0.389) \\
Complexity & -0.001 & 0.186 & -3.386** & 0.073 & -0.185 \\
 & (0.110) & (0.117) & (1.329) & (0.364) & (0.280) \\
TPV (t-1) & 0.560*** & 0.405** & 0.844*** & 0.277* & 0.270 \\
 & (0.164) & (0.173) & (0.301) & (0.163) & (0.190) \\
Number of destinations (log of) (t-1) & -0.055 & -0.087 & 0.441 & 0.245 & -0.134 \\
 & (0.274) & (0.314) & (0.332) & (0.305) & (0.317) \\
Firm's number of products (log of) (t-1) & -0.082 & -0.056 & 0.153 & 0.013 & -1.208*** \\
 & (0.110) & (0.118) & (0.567) & (0.078) & (0.362) \\
Constant & 10.079*** & 10.559*** & 9.192*** & 6.168*** & 10.263*** \\
 & (1.279) & (1.500) & (1.635) & (0.822) & (1.984) \\
Observations & 16,987 & 7,206 & 3,181 & 2,449 & 2,968 \\
\midrule
 \multicolumn{6}{c}{\textit{Estimation results for Peru}} \\
\midrule
Exports (log of) (t-1) & 0.566*** & 0.494*** & 0.751*** & 0.668*** & 0.542*** \\
 & (0.137) & (0.180) & (0.039) & (0.051) & (0.094) \\
Coreness (t-1) & 1.154* & 1.371 & 0.684*** & 0.930*** & 0.802 \\
 & (0.656) & (0.872) & (0.241) & (0.307) & (0.708) \\
Year 2020 & -0.314** & -0.322* & -0.171** & -0.259** & -0.699*** \\
 & (0.153) & (0.194) & (0.075) & (0.102) & (0.216) \\
Year 2020 * Coreness (t-1) & 0.108 & 0.119 & -0.137 & -0.112 & 0.145 \\
 & (0.191) & (0.236) & (0.109) & (0.137) & (0.402) \\
Complexity & -0.408*** & -0.562*** & 0.029 & -0.446* & -0.496* \\
 & (0.157) & (0.214) & (0.171) & (0.242) & (0.274) \\
TPV (t-1) & 0.075 & 0.035 & 0.136 & 0.118 & 0.279* \\
 & (0.078) & (0.089) & (0.109) & (0.115) & (0.149) \\
Number of destinations (log of) (t-1) & -0.392** & -0.397* & -0.166 & 0.485*** & 0.653 \\
 & (0.192) & (0.216) & (0.141) & (0.149) & (0.454) \\
Firm's number of products (log of) (t-1) & 0.050 & 0.052 & 0.096 & -0.187** & -0.484 \\
 & (0.089) & (0.101) & (0.076) & (0.091) & (0.307) \\
Constant & 7.491*** & 8.494*** & 4.001*** & 3.817*** & 6.401*** \\
 & (1.972) & (2.577) & (0.649) & (0.775) & (1.383) \\
Observations & 63,341 & 28,341 & 11,996 & 8,811 & 9,829 \\
\bottomrule
\multicolumn{6}{p{14cm}}{\textit{Notes:} Robust standard errors in parentheses. All estimations include fixed effect dummies for firms interacted with product HS chapter (first two digits of the HS code). Significance level: *** p$<$0.01, ** p$<$0.05, * p$<$0.10.}
\end{tabular}}
\end{center}
\end{table}

\clearpage
\newpage
\subsection*{Econometric estimations including single product firms}

\begin{table}[h!]
\begin{center}
\scriptsize 
\caption{Estimation results all countries pooled including single product firms.}
\label{tb:app_ppml_single}\vspace{3pt}
\resizebox{0.8\textwidth}{!}{\renewcommand{\arraystretch}{1.2}
\begin{tabular}{l ccccc} 
\toprule
Model &	(1)	&	(2)	&	(3)	&	(4)	&	(5)	\\
\midrule
 & Full sample & \multicolumn{4}{c}{Complexity Quartile}  \\
 \cmidrule(lr){2-2}
\cmidrule(lr){3-6} 
 &  & 1st & 2nd & 3rd & 4rd \\
\midrule
Exports (log of) (t-1) & 0.514*** & 0.459*** & 0.414*** & 0.692*** & 0.542*** \\
 & (0.048) & (0.064) & (0.091) & (0.030) & (0.054) \\
Coreness (t-1) & 1.604*** & 1.785*** & 2.073*** & 0.925*** & 1.889*** \\
 & (0.240) & (0.326) & (0.450) & (0.195) & (0.402) \\
Year 2020 & -0.245** & -0.307** & -0.291* & -0.207*** & -0.175 \\
 & (0.101) & (0.142) & (0.151) & (0.060) & (0.146) \\
Year 2020 * Coreness (t-1) & -0.027 & 0.044 & 0.122 & -0.038 & -0.279 \\
 & (0.119) & (0.158) & (0.192) & (0.106) & (0.183) \\
Complexity & -0.123* & -0.143 & -0.487 & -0.163 & -0.080 \\
 & (0.070) & (0.112) & (0.385) & (0.125) & (0.156) \\
TPV (t-1) & 0.102* & 0.058 & 0.323** & -0.040 & 0.200** \\
 & (0.060) & (0.068) & (0.129) & (0.074) & (0.088) \\
Number of destinations (log of) (t-1) & -0.087 & -0.031 & -0.118 & -0.078 & -0.107 \\
 & (0.099) & (0.109) & (0.136) & (0.118) & (0.169) \\
Firm's number of products (log of) (t-1) & -0.069 & -0.083 & 0.068 & 0.030 & -0.203 \\
 & (0.049) & (0.054) & (0.112) & (0.074) & (0.192) \\
Single product firm (t-1) & -0.205** & -0.163** & -0.992** & 0.286 & -0.379 \\
 & (0.082) & (0.073) & (0.424) & (0.191) & (0.235) \\
Constant & 7.995*** & 8.890*** & 8.259*** & 4.505*** & 6.883*** \\
 & (0.752) & (0.987) & (1.112) & (0.455) & (0.650) \\
Observations & 143,586 & 60,426 & 27,713 & 23,018 & 23,703 \\
Firm-HS2 FE & Yes & Yes & Yes & Yes & Yes \\
\bottomrule
\multicolumn{6}{p{14cm}}{\textit{Notes:} Robust standard errors in parentheses. Significance level: *** p$<$0.01, ** p$<$0.05, * p$<$0.10.}
\end{tabular}}
\end{center}
\end{table}

\end{document}